\documentclass[useAMS,usenatbib]{mn2e}

\include{journals}
\usepackage{times}
\usepackage{multirow} 
\usepackage{graphicx}
\usepackage{subfigure}
\usepackage{natbib}
\bibpunct{(}{)}{;}{a}{}{,} % to follow the A&A style
\usepackage{txfonts}
\newcommand{\xmm}{{\it XMM-Newton}}
\newcommand{\ace}{{\it ACE}}
\newcommand{\suzaku}{{\it Suzaku}}
\newcommand{\chandra}{{\it Chandra}}
\newcommand{\wind}{{\it Wind}}

\mathchardef\mhyphen="2D
\def\ourc{Obs301}
\def\prevf{Obs101}
\def\prevs{Obs201}
\def\erad{$\rm R_{E}$}

\title[A CME as seen by \xmm]{A high charge state Coronal Mass
  Ejection seen through solar wind charge exchange emission as
  detected by \xmm} \author[J. A. Carter, S. Sembay and
A. M. Read]{J. A. Carter$^{1}$\thanks{E-mail: jac48@star.le.ac.uk
    (JAC); sfs5@star.le.ac.uk (SS); amr30@star.le.ac.uk
    (AMR)}, S. Sembay$^{1}$ and A. M. Read$^{1}$\\
  $^{1}$The University of Leicester, Leicester, UK}
\begin{document}

\date{Accepted . Received ; in original form }

\pagerange{\pageref{firstpage}--\pageref{lastpage}} \pubyear{2009}

\maketitle

\label{firstpage}

\begin{abstract}  
  We present the analysis of an observation by \xmm\ that exhibits
  strongly variable, low-energy diffuse X-ray line emission. We reason
  that this emission is due to localised solar wind charge exchange
  (SWCX), originating from a passing cloud of plasma associated with a
  Coronal Mass Ejection (CME) interacting with neutrals in the Earth's
  exosphere. This case of SWCX exhibits a much richer emission line
  spectrum in comparison with previous examples of geocoronal SWCX or
  in interplanetary space. We show that emission from O\,VIII is very
  prominent in the SWCX spectrum. The observed flux from oxygen ions
  of $\rm 18.9\,keV\,cm^{-2}\,s^{-1}\,sr^{-1}$ is consistent with SWCX
  resulting from a passing CME. Highly ionised silicon is also
  observed in the spectrum, and the presence of highly charged iron is
  an additional spectral indicator that we are observing emission from
  a CME. We argue that this is the same event detected by the solar
  wind monitors \ace\ and \wind\ which measured an intense increase in
  the solar wind flux due to a CME that had been released from the Sun
  two days previous to the \xmm\ observation.
\end{abstract}


\section{Introduction}\label{secintro}

We report on a case of extreme solar wind charge exchange (SWCX)
detected by X-ray signatures within an observation taken with the
\xmm\ telescope \citep{jansen}. SWCX occurs when an ion in the solar
wind interacts with a neutral atom and the ion gains an electron into
an excited state. If the ion is in a sufficiently high charge state,
single or multiple X-ray or UV photons are emitted in the subsequent
relaxation of the electron. In the case of geocoronal SWCX, i.e. in
the vicinity of the Earth, the neutrals involved in the charge
exchange interaction are exospheric hydrogen. A SWCX spectrum is
characterised by emission lines from ions such as C, N, O, Mg and Fe,
without a continuum component.

For the general user of the \xmm\ observatory, the short-term
variations from geocoronal SWCX \citep{snowden2004, carter2008}, or
the longer term variations that occur from SWCX interactions in
interplanetary space \citep{smith2005} or at the heliosheath
\citep{cravens2000, koutroumpa2007}, can produce a significant diffuse
signal below $\sim 2$\,keV. This signal may constitute an additional
source of background when studying astrophysical sources, which must
be characterised before being removed from an astrophysical data
set. However, geocoronal SWCX provides important information
concerning the constituents of the solar wind, supplementing or
contributing additional information to that received by upstream solar
wind monitors. Solar wind compositional signatures such as abundances
and abundance ratios are indicative of the conditions prevailing near
the Sun during the acceleration of the solar wind, a process that is
not fully understood \citep[and references
therein]{richardson2004}. Geocoronal SWCX has been proposed as a
mechanism to image the magnetosheath around the Earth
\citep{collier2005, collierlxo}, leading to an increase in
understanding of transport processes within the plasma in and near the
bow shock. This case of SWCX therefore holds interest for both the
astronomical and solar-terrestrial communities.

The \xmm\ observation under study in this paper was made on the
$21^{st}$ of October, 2001. We present the analysis of this
observation and reason that the unusual X-ray signatures seen are due
to a Coronal Mass Ejection (CME) that was recorded on $19^{th}$ of
October 2001 by the Solar and Heliospheric Observatory (SOHO)
\citep{domingo1995} and which subsequently passed by the Earth. CMEs
involve an ejection of high density plasma with characteristics
different to that of the ambient solar wind; for example unusually
high Fe charge states or enhanced alpha particle to proton ratios
\citep{zurbuchen}. CMEs may pass by the Earth, depending on their
location of origin in the solar corona and passage through
interplanetary space. The absolute frequency of CMEs increases around
solar maximum, although at solar minimum, CMEs occur approximately
weekly. The event under analysis in this paper occurred close to solar
maximum in 2001. We use additional data from both the Advanced
Composition Explorer (\ace, \citet{stone1998}) and the \wind\
\citep{acuna1995} spacecraft to support our argument. We also analyse
\xmm\ observations before and after our case observation and the
nearest \chandra\ observation in time to this period.

The $21^{st}$ of October 2001 event was first identified as a
particularly noteworthy observation during a systematic search for
SWCX within the \xmm\ public archive as described in
\citet{carter2008}. That study performed an analysis of 187
observations of the EPIC-MOS instruments \citep{turnershort} taken in
full-frame imaging mode to search for variable diffuse emission in an
energy band concentrating on the SWCX indicators of O\,VII and O\,VIII
emission between 0.5\,keV and 0.7\,keV.

This observation proved to be the most spectrally rich example of SWCX
found within the sample, and its possible association with a known CME
warranted a more detailed study. In fact, as we shall show, the
dominant diffuse component in the entire observation was due to X-rays
from SWCX. In this paper we perform a more detailed spectral analysis
than reported in the survey paper of \citet{carter2008}. We extend our
analysis to include the EPIC-pn which was also in full frame mode for
this observation.

The paper is organised as follows. In section \ref{secmulti} we
describe several relevant \xmm\ observations of the same target as the
$21^{st}$ of October event and a \chandra\ observation taken around
the same time. Section \ref{secnewdata} contains a spectral analysis
of the \xmm\ data and in Section \ref{secview} we discuss the viewing
geometry and orientation of \xmm\ during the observation. We end the
paper with Section \ref{secdiscuss} containing our discussions and
conclusions.

\section{XMM-Newton and Chandra pointings}\label{secmulti}

\begin{table}
  \caption{\xmm\ observations from October 2001, towards right ascension and
    declination of 08h\,49m\,06s, +$44^{\circ}$\,51'\,24''. We state
    the orbital revolution number (one orbit takes 48 hours), instrument 
    and exposure identifiers as explained in the text. The start and stop 
    times are given in the \xmm\ time system which is the number of seconds 
    since the start of 1999. All observations were taken with the medium filter
    and in \textit{full-frame} mode.}
\begin{tabular}{llllll}
\hline
Obs id.    & Rev  & Inst & ExpID & Start & Stop \\  
           &  &   &  & ($\times$$10^{8}$\,s) & ($\times$$10^{8}$\,s) \\  
\hline
0085150101 & 0339 & MOS1  & S002  & 1.195187 & 1.195674 \\
           &      & MOS2  & S003  & 1.195185 & 1.195674 \\
           &      & pn    & S001  & 1.195211 & 1.195671 \\
\hline
0085150201 & 0342 & MOS1  & S002  & 1.200241 & 1.200703 \\
           &      &       & U002  & 1.200739 & 1.200743 \\
           &      & MOS2  & S003  & 1.200241 & 1.200704 \\
           &      &       & U002  & 1.200742 & 1.200743 \\
           &      & pn    & S001  & 1.200265 & 1.200746 \\
\hline
0085150301 & 0342 & MOS1  & S002  & 1.200781 & 1.200800 \\
           &      &       & U002  & 1.200837 & 1.200838 \\
           &      &       & U003  & 1.200865 & 1.201288 \\
           &      & MOS2  & S003  & 1.200781 & 1.200801 \\
           &      &       & U002  & 1.200837 & 1.200839 \\
           &      &       & U003  & 1.200865 & 1.201288 \\
           &      & pn    & S001  & 1.200805 & 1.201286 \\
\hline
\end{tabular}
\label{tabxmm}
\end{table}

The $21^{st}$ of October 2001 SWCX event was recorded in an \xmm\
observation of target 1Lynx.3A\_SE (right ascension 08h\,49m\,06s and
declination +$44^{\circ}$\,51'\,24''). This is a field that contains
no bright point or extended source emission. The Galactic column in
the direction of this field is low ($\rm 2.79 \times 10^{20} \
cm^{-2}$). Fortuitously there were two additional observations of the
same target field taken around 6 days and 15 hours previous to the
SWCX event and with substantially overlapping fields of view; the
pointing directions of these observations being offset by 1.4 and 2.9
arcminutes respectively which is small compared to the circular 30
arcminute field of view of \xmm.

The observations and their start and stop times for the EPIC
instruments are detailed in Table \ref{tabxmm}. The identification
numbers for these observations are 0085150101, 0085150201 and
0085150301 in the nomenclature of the \xmm\ science
archive\footnote{http://xmm.esac.esa.int/xsa/}. Henceforth they are
referred to as \prevf, \prevs\ and \ourc\ (the SWCX event)
respectively. Breaks during a single observation, noted using various
exposure identifiers, are due to the instruments being switched to a
safe, non-observational mode as a result of the extremely high
radiation environment that the satellite encountered during this
period. No other \xmm\ observation was performed between \prevs\ and
\ourc.

In \citet{carter2008} we extracted from our sample observations
lightcurves of the diffuse X-ray signal in an energy band that could
potentially contain O\,VII and O\,VIII line emission from SWCX
(0.5\,keV to 0.7\,keV) and also for comparison a non-SWCX continuum at
higher energies (2.5\,keV to 5.0\,keV). The latter band can be used to
exclude cases where variability observed in the low energy band is due
to a variable particle background. The observations were essentially a
random sample from a set of publicly available MOS full-frame
imaging mode observations that passed the criteria of having good
exposure times and relatively weak sources in the
field-of-view. \prevs\ was also included in the sample and, unlike
\ourc, showed no evidence for a variable SWCX component to the
observed diffuse emission.

In addition, we searched for \chandra\ \citep{weisskopf2000}
observations from the \chandra\
archive\footnote{http://cxc.harvard.edu/cda/} during the period of the
\ourc\ event, but unfortunately there was no simultaneous
coverage. The closest observation, (number \textit{2365}, instrument
ACIS-I, target 1RXSJ161411.3-630657), began towards the end of
\prevs\, but was stopped well before the start of \ourc\ due to the
high radiation environment also experienced by \chandra\ at the time.

The \xmm\ observation immediately after the \ourc\ observation was
very heavily radiation contaminated and extremely short so was
excluded from further analysis. This was followed by several
observations in a closed calibration mode (CALCLOSED). The observation
after these CALCLOSED observations (observation 0083000101, target
B3\,0731+438) was also in full-frame mode for each of the EPIC
instruments.

\begin{figure*}
  \centering
  \vspace{1.0cm}
  \includegraphics[width=0.75\textwidth, angle=90, bb=80 300 550 470]{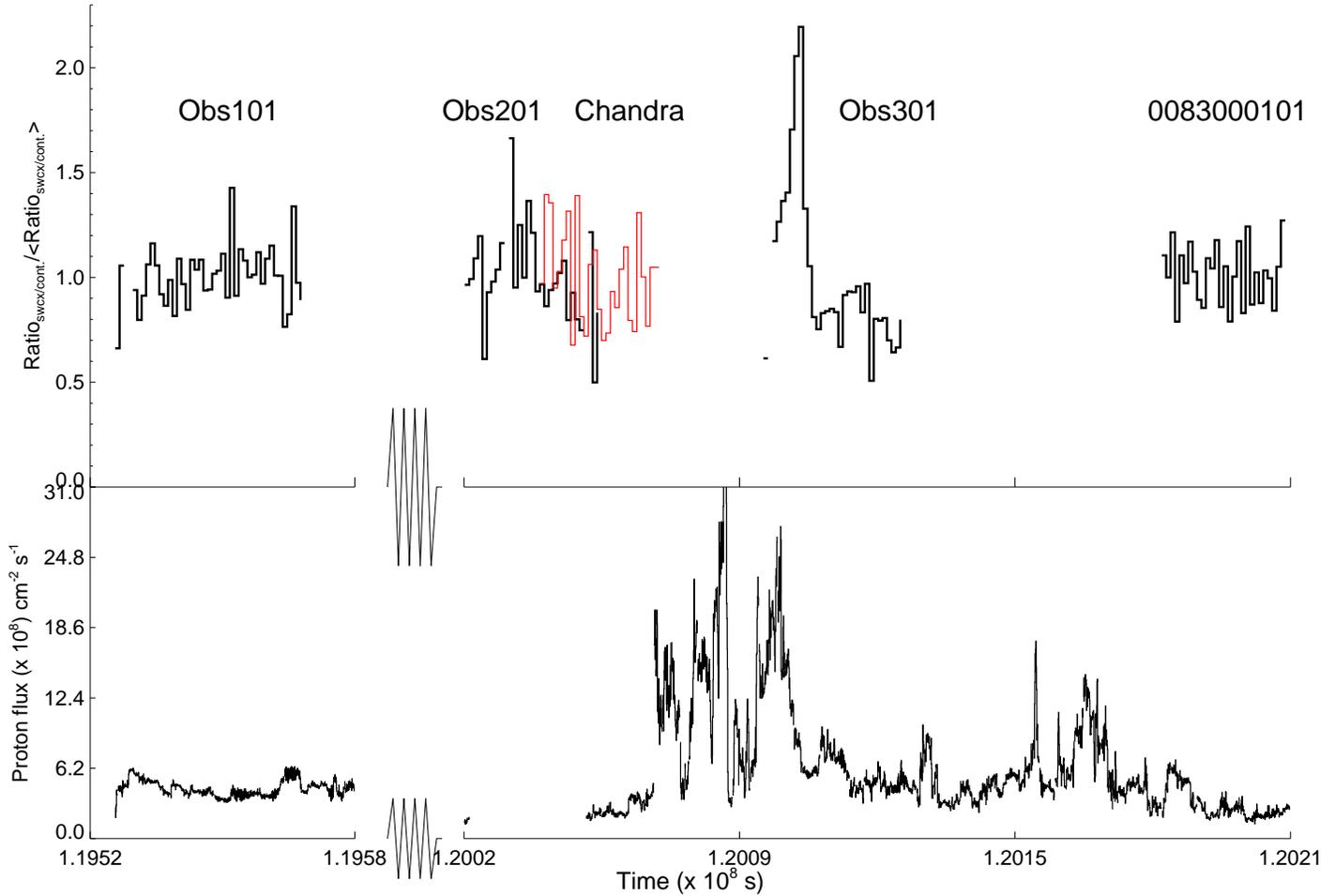}
  \caption{Upper panel: The ratio of the (0.5\,keV to 0.7\,keV, oxygen
    emission line band) and (2.5\,keV to 5.0\,keV, continuum band)
    lightcurves, scaled to the mean of this ratio for each
    observation, for observations \prevf, \prevs, and \ourc\ along
    with the next \xmm\ imaging mode observation that was sufficiently
    long ($>$4\,ks) to be processed (observation id. 0083000101). The
    same ratio scaling is used for the \chandra\ data, shown in
    red. Lower panel: The solar wind proton flux, taken from the \ace\
    SWEPAM instrument.}
     \label{figcomblc}
\end{figure*}

In Figure \ref{figcomblc} we plot an illustration of the X-ray
activity as seen by these observations. We have plotted the ratio of
the diffuse ``Oxygen'' line band (0.5\,keV to 0.7\,keV) count rate to
continuum band (2.5\,keV to 5.0\,keV) count rate, normalised by the
mean of this ratio for each observation. All \xmm\ data have been
filtered as described in detail in \citet{carter2008} and later in
this paper (Section~\ref{secnewdata}). For the \chandra\ observation
we downloaded the ACIS \textit{level 2} event file and extracted the
counts from a large region (radius 0.06 degrees) centred on chip 3. By
inspection the \chandra\ target is an extended and presumably
non-variable source. The \xmm\ \ourc\ observation is the only
observation with evidence for a variable diffuse signal in the low
energy line band that is not correlated with variations in the higher
energy band.

Using the same time axis we also show the solar proton flux, as
recorded by \ace\ at the sunward L1 Lagrangian point, approximately 200
earth radii (\erad) from the Earth. The data are Level 2 64
second-averaged products from the Solar Wind Electron Proton Alpha
Monitor (SWEPAM) \citep{mccomas1998}
archive\footnote{http://www.srl.caltech.edu/ACE/ASC/}. One can see a
dramatic rise in the solar proton level between \prevs\ and \ourc. It
has been shown by \citet{wang2005} (see also this paper,
Section~\ref{secdiscuss}) that the rise in activity recorded by \ace\
at this epoch was due to the $19^{th}$ October 2001 CME. A rise in
the \xmm\ ratio is seen shortly afterwards. 
%%%% This following paragraph is part of the previous section
\xmm, however, does not necessarily sample the same solar wind as that
measured by \ace\ due to their spatial separation. Other geometric
factors play a role, such as the pointing direction of \xmm, which we
discuss in more detail in Section~\ref{secview}. In addition, the
proton flux is only a proxy to the behaviour of the heavy ion content
of the solar wind. The expected X-ray emission from SWCX depends on
the abundances, cross-sections and velocities of the ions involved,
but we must also factor in the integration along the line of sight
that \xmm\ takes through the interacting region, in the vicinity of
the Earth.
%%%%%%%%%%%%%

Nevertheless, as we shall argue in Section~\ref{secdiscuss}, the SWCX
event seen in the \xmm\ \ourc\ data is sampling the same CME event
recorded by \ace\ shortly before. In the following section we will
concentrate our X-ray analysis on the two \xmm\ observations, \prevf\
and \ourc. \prevf\ is useful because it allows us to unambiguously
determine the non-variable diffuse X-ray background in the direction of
\ourc.

\section{XMM-Newton data analysis}\label{secnewdata}

The Science Analysis System (SAS) software (version 8.0.0;
\textit{http://xmm.esac.esa.int/external/xmm\_data\_analysis/}) was
used to process the raw data into calibrated event lists and extract
light curves, spectral products and instrument response files. The
instrument effective area files were calculated assuming the source
flux is extended with no intrinsic spatial structure. The SAS accesses
instrument calibration data in so-called current calibration files
(CCFs) which are generally updated separately from SAS release
versions. In this paper we used the latest public CCFs released as of
July 2009.

When creating products from the calibrated event files we applied the
following filter expressions in the nomenclature of the SAS;
(PATTERN$<$=12)\&\&(\#XMMEA\_EM) for the MOS and
(PATTERN==0)\&\&(FLAG==0) for the pn. The specified PATTERN filter
selects events within the whole X-ray pattern library for the MOS and
mono-pixels only for the pn; as our focus is on detecting line
emission below 2\,keV, this restriction optimises the energy
resolution of the pn with little loss of sensitivity in the energy
range of interest. For these event class selections, the energy
resolution of the pn is $\sim 70$\,eV (FWHM) at 1\,keV compared with
$\sim 60$\,eV in the MOS. The filter \#XMMEA\_EM removes events from
the MOS that are from regions of known bright pixels or columns or
near CCD boundaries (which tend to be noisy). The equivalent flag for
the pn, \#XMMEA\_EP, did not remove some residual noisy pixels, but
these were removed when we used the more conservative FLAG==0. This
flag also masks out events from adjacent regions to noisy pixels.

In addition we selected only events within a radius of 11.7
arcminutes, centred on a communal sky position such that the
extraction region of all three cameras was covered by active silicon,
barring inter-CCD gaps.

A spectral analysis of the SWCX emission component in \ourc\ requires
us to identify and account for each of the sources of X-rays that
contribute to the combined signal across the field of view. A detailed
description of the various components which constitute the \xmm\ EPIC
background is given in \citet{carter2007}. In the following sections
we describe how each of these components, plus the contribution from
resolved point sources, is either subtracted from our data, or
modelled within our spectral fitting. We have used Version 12.5.0 of
the
XSPEC\footnote{http://heasarc.gsfc.nasa.gov/docs/xanadu/xspec/index.html}
X-ray spectral fitting package to perform this analysis.

\subsection{Resolved point sources}\label{secpointsources}

\begin{figure*}
  \centering
  \includegraphics[width=0.3\textwidth, angle=270]{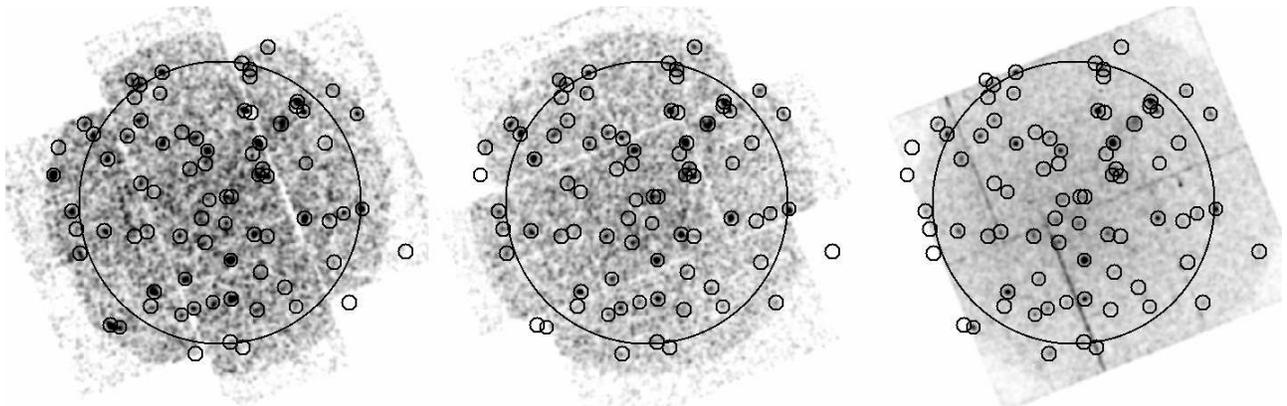}
  \caption{Images from each of the EPIC cameras for \ourc\ (left to
    right); MOS1, MOS2 and pn, in the energy range 0.3\,keV to
    2.0\,keV. The small black circles indicate the point sources
    removed from analysis, and the large black circle indicates the
    spatial extraction region used for spectral analysis.}
     \label{figremsources}
\end{figure*}

Resolved point sources can potentially contribute a spectrally
variable signal and thus need to be removed as far as possible from
our datasets. To do this we used the source lists that are
automatically produced for the processing of the 2XMM catalogue and
are available as a product within the \xmm\ archive. These lists were
used to create an exclusive filter that removed events from the
calibrated event lists out to a radius of 35 arcseconds ($\sim 90$\%
of the on-axis point spread function) from the centre of each
source. Using the source count rates within the source lists, we
estimate that the total residual resolved source count rate (0.2 to
2.0 keV) in the pn after cleaning would be around $\rm
0.04\,cts\,s^{-1}$ in \prevf. The background subtracted count rate in
the same energy band after source removal was $\rm 0.75 \ cts \
s^{-1}$, hence, residual sources contribute approximately 5\% of the
observed diffuse flux in this observation. In \ourc\ the diffuse flux
count rate is $\rm 1.8\,cts\,s^{-1}$ and the residual contribution
from resolved point sources is at around
2\%. Figure~\ref{figremsources} shows the cleaned images from each
camera from \ourc, with the positions of the point sources and
spectral extraction region marked, in the energy range 0.3\,keV to
2.0\,keV. There are 62 sources which overlap the source extraction
region.

\subsection{Soft proton contamination}

\begin{figure}
      \includegraphics[width=.475\textwidth,height=100mm]{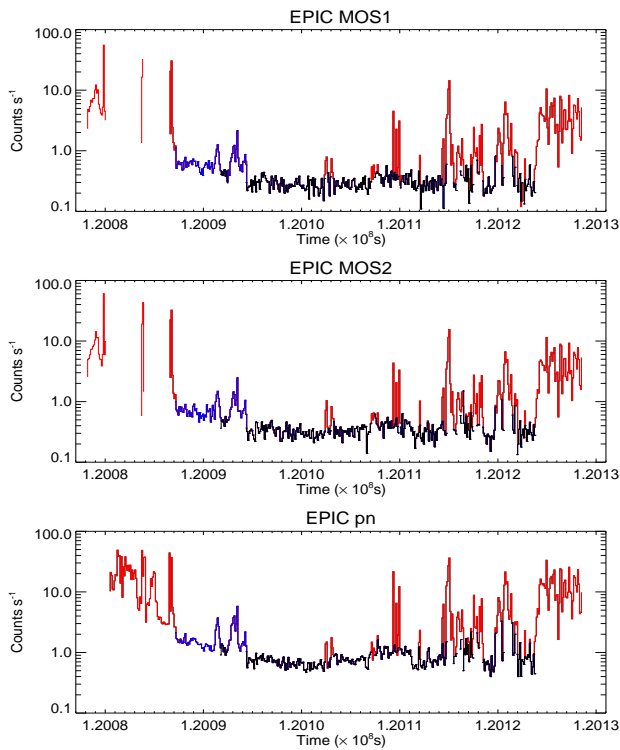}
      \caption{The 2.5\,keV to 8.5\,keV EPIC lightcurves of the
        diffuse signal from observation \ourc, incorporating all
        exposures. The variance in the signal is due to soft proton
        contamination. Bins marked in red indicate where data have
        been excluded from further analysis. Bins marked in black are
        the cleanest data used for spectral fitting of the integrated
        spectra. Bins in blue are used in addition for analysis of the
        lightcurve.}
      \label{figsoftproton}
\end{figure}

Soft protons can produce an extremely variable, spectrally smooth,
wide-band, whole field of view signal that can be the dominant source
of X-rays within an observation. The particles may be solar protons
accelerated by reconnection events in the vicinity of the Earth
\citep{lumb2002}. They are funnelled by the telescope mirrors onto the
detectors where they are absorbed; the signals produced by individual
events are indistinguishable from X-rays. Figure~\ref{figsoftproton}
shows the \ourc\ 2.5\,keV to 8.5\,keV light curves of the three EPIC
cameras (after point-source removal) binned in 100\,s intervals. The
dataset is extremely contaminated by soft proton flares and show data
gaps in the MOS when the instrument was switched to a safe mode. The
pn instrument takes longer to setup for a given observation therefore
the pn lightcurve starts about 40 minutes after the MOS.

Data cleaning schemes for flares include excluding bins whose count
rate in a high energy band exceeds a set threshold or excluding bins
which are more than a set value of sigma from the mean value of the
light curve. The latter method is that employed by the publicly
available Extended Source Analysis Software
(ESAS)\footnote{http://heasarc.gsfc.nasa.gov/docs/xmm/xmmhp\_xmmesas.html}
package, as described in \citet{snowden2008}. At the time of writing
the ESAS package is only directly applicable to the MOS data, however,
we used the package to derive Good Time Interval (GTI) files for MOS1
and MOS2 and then merged them into a single file which proved to be
sufficiently accurate in identifying periods of proton flaring in the
pn data.

The data bins accepted by the single GTI file are shown in
Figure~\ref{figsoftproton} in black. We have used this data selection
for the integrated spectra as described in Section~\ref{secswcxspec}. We
also included the data for the period marked in blue in our analysis
of the SWCX lightcurve because we wished to try and establish the
start of the period of SWCX enhancement.

Soft protons can have a low temporal variance which is often difficult
to detect via an analysis of the lightcurve, although in this case it
can be seen that there appears to be a slowly varying signal in the
residual GTI periods. This was confirmed as being due to soft protons
via our spectral analysis.

The spectral signature of soft protons has been studied by
\citet{kuntz2008}. These events produce a featureless {\it power-law}
spectrum unmodified by the detector response. There is generally a
correlation between spectral hardness and intensity and the spectral
slope can show a break with a steepening at higher energies. The
component can be modelled within a typical XSPEC analysis session by
folding a power-law (or broken power-law) through a diagonal response
matrix (i.e. one constructed to have a response value of 1.0 on the
diagonal elements and zero elsewhere). XSPEC Version 12 has the
functionality to enable several model spectra to be convolved with
distinct responses and then added together into a combined model which
can be compared with the data. This combined model refers to the model
of the diffuse X-ray emission, which incorporates various background
components convolved with the instrument response and the soft proton
model convolved with the diagonal response.

\subsection{Cosmic-ray particle background (CPB)}\label{secCPB}

High energy cosmic rays produce background either directly within the
CCDs or via fluorescence within the spacecraft material surrounding
the detectors. This source of background is relatively stable in
spectral shape and intensity. The CPB contribution to a given
observation can best be estimated by deriving the spectrum from EPIC
data taken when the instruments are in the filter wheel closed (FWC)
configuration (i.e not open to the sky). In this configuration the
observed signal consists of the CPB plus detector noise
(Section~\ref{secdetnoise}). The \xmm\ Background Working Group
(BGWG) maintain co-added event files of FWC data (total exposures of
around 700 ksec in each MOS and 300 ksec in the pn) on their public
web site
\footnote{http://xmm.vilspa.esa.es/external/xmm\_sw\_cal/background}.

The CPB spectrum varies across the field of view of each instrument,
therefore, it is necessary to extract the spectra from the identical
regions to those that define the source spectrum. The spectra so
extracted constituted the background files in our XSPEC fits.

It is not unusual for the derived CPB spectrum to require some small
amount of re-scaling for a given observation. This can be done by
comparing the observed high energy count rate in the source and CPB
spectra above an energy where the contribution from components other
than the CPB in the source spectrum is expected to be negligible.
Naturally, the source dataset must be clean of soft proton
contamination before a simple scaling of the CPB can be made. As this
was the case in \prevf\ and as our model of the diffuse X-ray sky
predicted a relatively negligible contribution to the observed count
rate in the energy range 7.75\,keV to 12.0\,keV, we used this band in
all three EPIC instruments to derive scaling factors for the CPB of
1.26, 1.11 and 1.08 respectively in the pn, MOS1 and MOS2 (i.e. the
observed FWC CPB count rate was greater by these factors than observed
in our source observation). Such factors are not uncommon
\citep{deluca2004}.

Because our SWCX dominated observation, \ourc, was contaminated by
residual soft protons throughout the observation and therefore had a
strong contribution from this component at high energies, we were
unable to apply the same procedure as for \prevf\ to subtract the CPB.
We assumed therefore that the scaling factors for the CPB derived from
\prevf\ would be appropriate for this observation; a reasonable
assumption given that the observations are only 6 days apart.

\subsection{Detector noise}\label{secdetnoise}

Detector noise consists of persistent and variable components, in both
a temporal and spatial sense. Persistent noise occurs from thermal
Poisson processes in each CCD pixel which creates events with
sufficient charge to appear above the detector threshold. This
component is essentially fixed. It is one component of the signal
within filter wheel closed datasets (see Section \ref{secCPB}) and it
is therefore fairly straightforward to subtract from the total signal.

Variable components are primarily due to pixels damaged by
radiation. Pixels so bright that they can cause the event rate to
exceed the instrument telemetry limit are blocked on board. Other
defective areas are recognised by the SAS and events from these
regions are flagged (see Section \ref{secnewdata}), enabling them to
be excluded or included depending on the requirements of the analysis.

There are other components, however, that are not so amenable to
subtraction via event flagging. CCD5 of the MOS1 detector showed an
elevated background across the whole chip characterised by a continuum
of spurious events with energies up to $\sim 1$\,keV. Our solution was
simply to remove this CCD from our analysis. This type of noise signal
is variable from observation to observation and other CCDs can also
show a similar behaviour with around $\sim 20$\% of observations
affected at some level \citep{kuntz2008}.  The physical cause of the
noise is unknown at present.

\subsection{Diffuse Galactic and extragalactic emission}\label{skyspec}

\prevf\ allowed us to independently derive the contribution from the
diffuse Galactic and extragalactic emission components in the
direction of \ourc. The target area has a galactic longitude and
latitude of ($176^{\circ}$, $40^{\circ}$) so is well away from the
plane of the Galaxy and Galactic centre. Having removed the resolved
point sources and flare cleaned the data, \prevf\ appeared by
inspection to contain no evidence for significant residual soft
protons or a time varying SWCX component. We therefore assumed that
the resultant diffuse emission comes from non-varying sources and that
the spectrum and intensity of this component could be fixed and
applied to observation \ourc.

Following previous authors (e.g. \citet{galeazzi}) we have modelled
the \prevf\ diffuse photon emission with a three component
description.  The first component is a constant un-absorbed plasma
representing emission from the local hot bubble and a possible
contribution from SWCX emission at the boundary of the heliosphere
\citep{robertson2003a, koutroumpa2007}. Any contribution from the
heliospheric SWCX we assume to be essentially constant over the 6 days
between the observations. The second component is an absorbed plasma
representing emission from the Galactic halo. We used the APEC
\citep{smith2001} model within XSPEC to model the plasma components
although the commonly used alternative Mekal and Raymond-Smith models
gave statistically identical results.  The third component, also
absorbed by the same line of sight material, is a power law
representing the unresolved extragalactic X-ray background from point
sources. For the absorption, we have used the {\it phabs} model within
XSPEC. The element abundances in the absorption and emission models
used were those set by the {\it wilm} table \citep{wilms2000}. The
value of $\rm N_{H}$ was fixed at the Galactic line of sight value of
$\rm 2.79 \times 10^{20} \ cm^{2}$ and derived using the nH tool
available from the HEASARC \footnote{http://heasarc.gsfc.nasa.gov}.

We first independently fit the model to each of the CPB background
subtracted \prevf\ spectra from the pn, MOS1 and MOS2. However, after
several analysis iterations we found that we could achieve an
acceptable fit to the data by jointly fitting the model to all three
cameras allowing only the global normalisation of the entire model to
vary between the cameras. Table~\ref{tabskymodel} lists the derived
model parameters and component fluxes. The fluxes from each camera are
consistent with each other at the 90\% confidence level, although the
MOS returns values $\sim 20$\% lower than the pn.

\begin{figure}
  \centering
  \includegraphics[width=0.475\textwidth, bb=55 360 550 695]{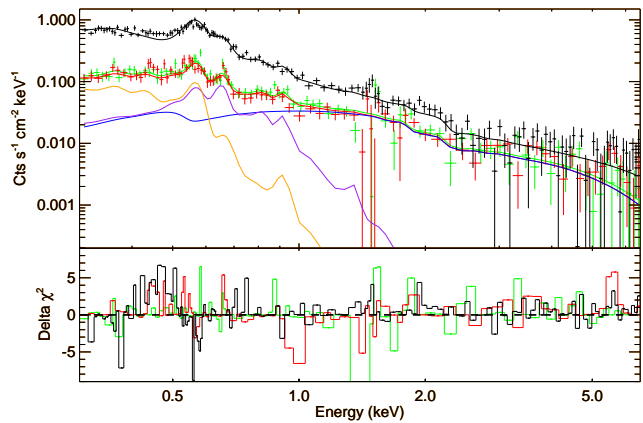}
  \caption{Background subtracted pn (black), MOS1 (red) and MOS2
    (green) spectra from \prevf\ compared with a model of the diffuse
    sky emission. Also shown are the three components of the model,
    the un-absorbed plasma (orange), the absorbed plasma (purple) and
    absorbed power-law continuum (blue). For clarity, only the
    individual model components for MOS1 are shown. The lower panel
    shows the deviation of the data from the model.}
     \label{fig101sp}
\end{figure}

Figure~\ref{fig101sp} shows the data compared to the best-fit model in
each of the three cameras. When fitting the data we excluded the
energy range 1.35\,keV to 1.9\,keV because, as can been seen in the
figure, the strong instrumental Al\,$\rm K_{\alpha}$ and Si\,$\rm
K_{\alpha}$ lines can produce large residuals at these energies after
background subtraction. There is also a broad residual in the pn fit
at 0.45\,keV whose strength is not sensitive to the particulars of
which abundance table or plasma model within XSPEC is selected. MOS1
has a similar, although narrower feature, whereas MOS2 does not. The
strength of the feature is not sufficient to have a significant
bearing on our analysis of the SWCX signal within \ourc.

\begin{table}
\centering
\caption{Sky model parameters in the direction of \ourc\ derived from 
  an analysis of \prevf. Quoted errors are 90\% confidence for one interesting parameter. Fluxes are observed values in the energy range 0.2\,keV to 10\,keV in units of $\rm 10^{-8}\,ergs\,cm^{-2}\,s^{-1}\,sr^{-1}$.}
\begin{tabular}{|ccc|}
\hline
\multicolumn{3}{|c|}{Diffuse Sky Background Model} \\
\hline
\multicolumn{2}{|c}{Reduced $\chi^{2}$ / Degrees of Freedom } & 0.99 / 1008 \\
\hline
Component & Parameter/Flux & Value (Error) \\
\hline
Unabs. Plasma & Temperature (keV) & 0.11 (0.01) \\
  & MOS1 Flux & 2.04 (0.18) \\
  & MOS2 Flux & 1.99 (0.18) \\
  & pn Flux   & 2.42 (0.20) \\
  & & \\
Abs. Plasma & Temperature (keV) & 0.23 (0.02) \\
  & MOS1 Flux & 1.23 (0.18) \\
  & MOS2 Flux & 1.12 (0.18) \\
  & pn Flux   & 1.46 (0.21) \\
  & & \\
Abs. CXRB & Photon Index & 1.44 (0.12) \\
  & MOS1 Flux & 5.31 (0.56) \\
  & MOS2 Flux & 5.18 (0.55) \\
  & pn Flux   & 6.29 (0.63) \\
\hline
\end{tabular}
\label{tabskymodel}
\end{table}

\subsection{Solar Wind Charge Exchange X-ray Emission}\label{secswcxspec}

\begin{figure*}
  \centering
  \includegraphics[width=0.85\textwidth, bb=55 30 515 700]{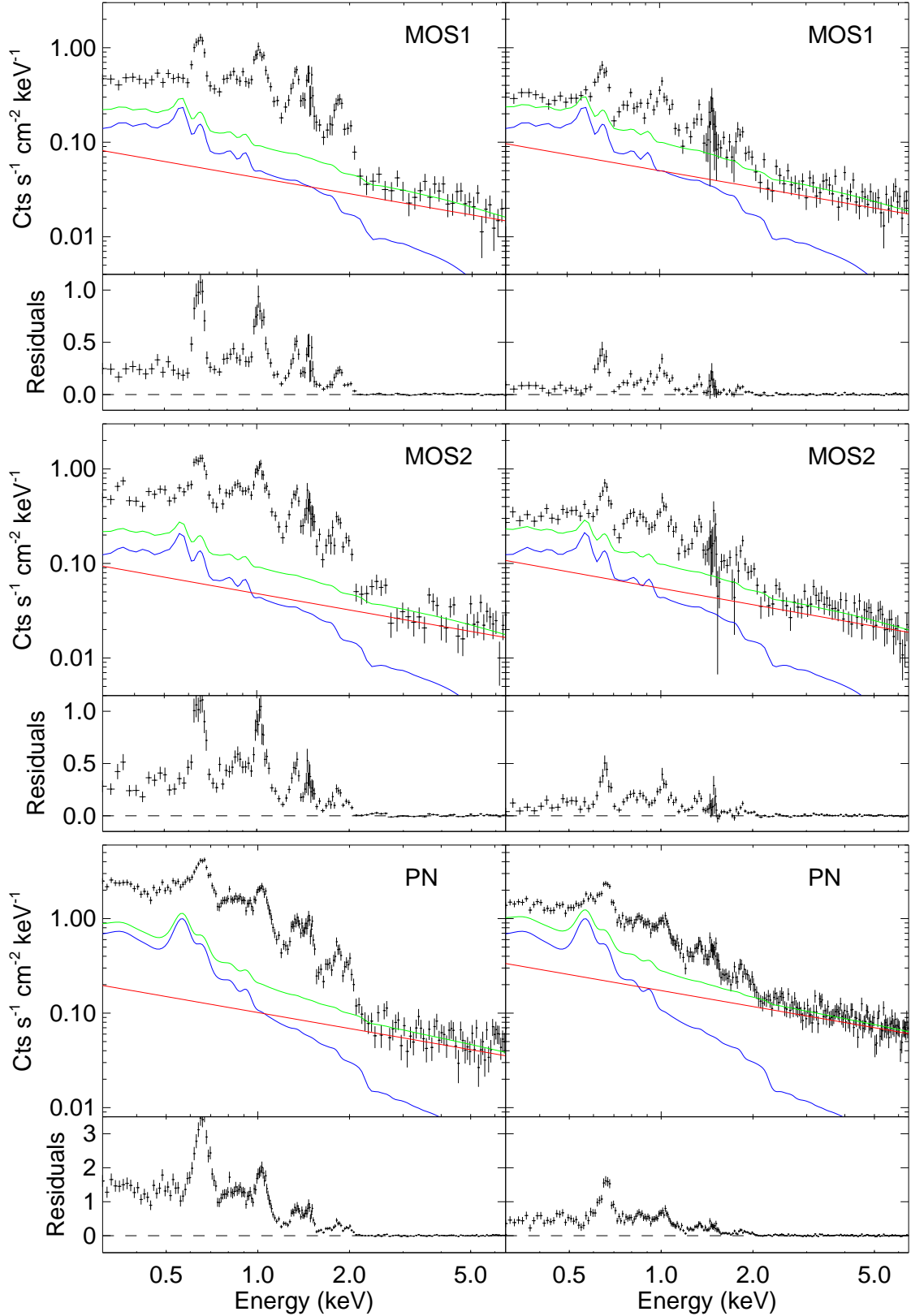}
  \caption{\ourc\ background-subtracted spectra integrated from $\rm T
    = 1.200918 \times 10^{8}\,s \ to \ 1.201035 \times 10^{8}\,s$ (left
    panels) and from $\rm T = 1.201035 \times 10^{8}\,s \ to \ 1.201230
    \times 10^{8}\,s$ (right panels). Contributions are shown from soft
    protons (red), the non-variable diffuse sky background (blue), and
    these components combined (green). The variable excess flux, seen
    in the lower panels, is due to SWCX.}
     \label{figsplit}
\end{figure*}

The lightcurve of \ourc\ in Figure~\ref{figcomblc} suggests the soft
band flux has a {\it flare} and a {\it quiescent} period. Background
subtracted spectra integrated over these intervals (extracted from the
soft proton flare-cleaned data) are shown in Figure~\ref{figsplit}. In
addition we show the diffuse sky model folded through the instrument
response and the strong residual soft proton component. The soft
protons are modelled with a single power law spectrum fit by initially
restricting the spectral fit to the range 2.5\,keV to 6.5\,keV. The
extrapolation of the power law shows that the soft proton component is
comparable to or weaker than the non-variable diffuse sky component
below $\sim 1$\,keV, and may be weaker still because, as previously
discussed, soft protons often display a spectral break. For this
reason, the strength of the residual SWCX component at low energies
may be underestimated by a few percent.

It is evident from Figure~\ref{figsplit} that {\em both} the flare and
quiescent periods show an excess flux above the combined diffuse sky
flux and soft proton contribution which is spectrally distinct from
these components. The excess clearly contains emission lines and is
variable which are both signatures of SWCX. The most prominent lines
are at 0.65\,keV and 1.02\,keV which we identify with O\,VIII and
Ne\,X. Laboratory measurements and theoretical calculations of SWCX
emission (e.g. \citet{wargelin2008}) indicate that emission spectra
will contain multiple lines from a variety of ions and their
transitions, most of which will be unresolved in the EPIC instruments,
given the limits of the energy resolution of the detectors.

Our spectral model of the SWCX excess was built up from a series of
zero width Gaussians with energies fixed at known X-ray emission
transitions from likely solar wind ions. For CV, CVI, NVI, NVII,
O\,VII and O\,VIII we have used the theoretical model by
\citet{bodewits2007b} (Table 8.2), who has calculated the relative
emission cross-sections of these bare and H-like ions (their state
before electron capture) in collision with atomic hydrogen for a
variety of solar wind velocities. We have used the tabulated values
for a velocity of $\rm 600 \ km \ s^{-1}$ which is close to the
velocity measured by ACE at this epoch. Our model fitting allowed the
six normalisations of the principal transition from each of these ions
to be free, but constrained the normalisations of the weaker
transitions to the ratios predicted by the Bodewits' model. In all,
these ions contributed 33 lines between 0.299\,keV and 0.849\,keV.

At higher energies we have taken a more empirical approach by adding
sufficient lines at known transition energies to characterise the bulk
of the residual excess emission. This will be an incomplete list due
to the multiple transitions expected from
Fe. Table~\ref{tabmodelparams} lists the principal transitions we have
included in the SWCX model.

We have fit our combined model (containing the SWCX, sky background
and soft proton components) jointly to the integrated spectra from
each of the EPIC cameras. The free parameters in the fit are the
normalisations of the principal ions in the SWCX model plus global
normalisations applied to the individual MOS spectra (the pn global
normalisation was fixed at 1.0).

In Table~\ref{tabmodelparams} we list the flux of the O\,VIII line at
0.65\,keV and the ratio of the fluxes for each of the other ions to
O\,VIII, and the total flux of the SWCX model. As with our analysis of
the diffuse sky background in \prevf, the broad-band MOS fluxes are
lower than that measured by the pn by a similar factor.

In a study of the inter-calibration of point sources from the 2XMM
catalogue, \citet{mateos2009} found the reverse trend; on average the
MOS cameras register a higher flux than the pn by 7-9\% below 4.5
keV. We can only attribute the difference to some unknown calibration
uncertainty in the calculation of the effective area for point sources
compared with an extended region.

\begin{figure}
  \centering
  \includegraphics[width=0.475\textwidth,height=120mm, bb=60 30 515 700]{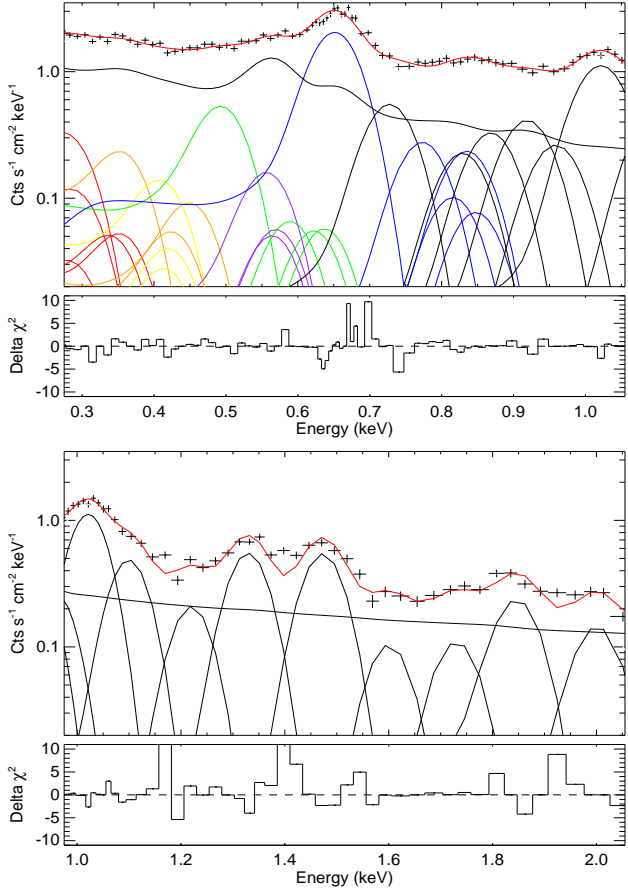}
  \caption{The SWCX spectral model fitted to the integrated background
    subtracted and flare-cleaned pn spectrum of \ourc. The top panel
    is the spectrum from 0.275\,keV to 1.055\,keV and the bottom panel
    is the spectrum from 0.975\,keV to 2.055\,keV. The sum of the
    non-variable sky and variable soft proton components is the
    continuous line in black. The lines due to C, N and O are colour
    coded; C\,V (red), C\,VI (orange), N\,VI (yellow), N\,VII (green),
    O\,VII (purple) and O\,VIII (blue). Heavier elements are in
    black. The residual at $\sim 1.4$\,keV may be due to incomplete
    background subtraction at the energy of the strong $\rm Al \
    K_{\alpha}$ instrumental line.}
     \label{figpnfit}
\end{figure}

Figure~\ref{figpnfit} shows the best fit SWCX spectral model (plus the
non-variable diffuse sky and variable soft proton components combined)
to the background-subtracted and flare-cleaned pn spectrum. The
non-Gaussian shape of the pn detector response (the MOS is similar;
the Gaussian shape is distorted by a low-energy shoulder) is evident
from the principal O\,VIII line.

\begin{figure}
  \centering
  \includegraphics[width=0.475\textwidth]{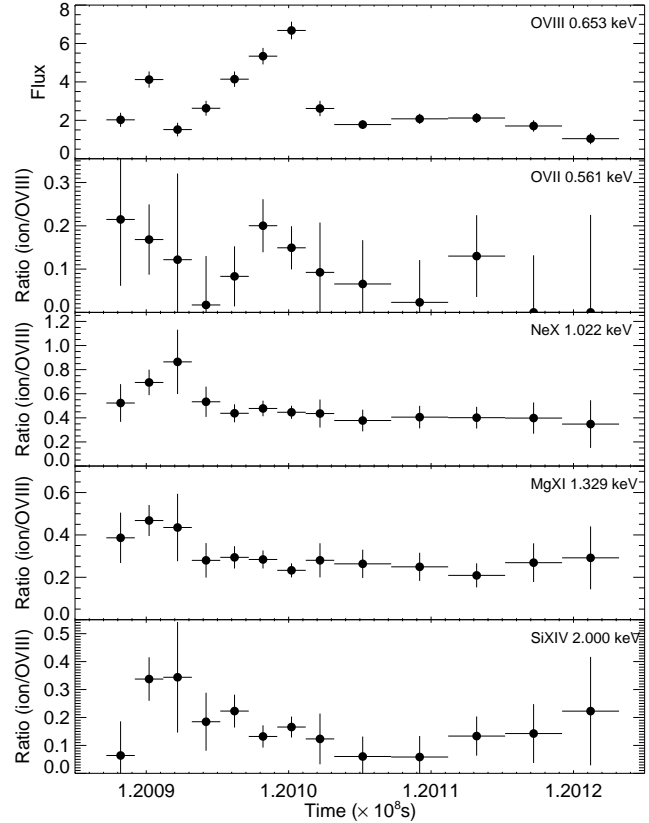}
  \caption{Lightcurves of selected ions from \ourc. The top panel is
    the OVIII flux in units of $\rm 10^{-8} \
    ergs\,cm^{-2}\,s^{-1}\,sr^{-1}$. Lower panels are the ratio of the
    flux of the selected ion to OVIII.}
     \label{figionflux}
\end{figure}

The temporal variation in the SWCX emission has been mapped by
extracting spectra in eight 2\,ks intervals followed by five 4\,ks
intervals. This covers the soft proton flare-cleaned period plus the
additional segment as shown in Figure~\ref{figsoftproton}. For each
interval, we show in Figure~\ref{figionflux} the fitted fluxes of the
O\,VIII (0.653\,keV) line and the ratio of the fluxes of OVII
(0.561\,keV), Ne\,X (1.022\,keV), Mg\,XI (1.329\, keV) and Si\,XIV
(2.000\,keV) to OVIII. There is little evidence for a significant
compositional change throughout the observation with the possible
exception of the second and third bins where the flux ratios of the
heavier ions are somewhat higher compared to the average.

Ion compositional data (level 2, hourly averaged) from the SWICS/SWIMS
instrument on board \ace\ \citep{gloeckler1998} are sparse for the
period of interest. \xmm\ therefore is able in this case to provide
supplementary abundance information where \ace, for data quality
reasons, cannot.

\begin{table}
\centering
\caption{Measured fluxes in the SWCX spectral model. Here we list only the principal transitions from C, N and O plus the additional selected transitions from Ne, Mg, Si and Fe. The figure for O\,VIII is the measured flux in units of $\rm 10^{-8} \ ergs \ cm^{-2} \ s^{-1} \ sr^{-1}$. The remainder of the table is the ratio of the measured flux for that ion to O\,VIII and the total flux of the SWCX model.}
\begin{tabular}{ccc}
\hline
Ion & Principal Energy (keV) & Ion Ratio / OVIII Flux \\
\hline
C\,V & 0.299 & 0.50(0.16) \\
C\,VI & 0.367 & 0.28(0.08) \\
N\,VI & 0.420 & 0.06(0.05) \\
N\,VII & 0.500 & 0.19(0.03) \\
O\,VII & 0.561 & 0.12(0.03) \\
\hline
O\,VIII & 0.653 & 2.70(0.09) \\
\hline
Fe\,XVII & 0.73 & 0.13(0.01) \\
Fe\,XVII & 0.82 & 0.05(0.02) \\
Fe\,XVIII & 0.87 & 0.10(0.03) \\
Fe\,XIX/Ne\,IX & 0.92 & 0.14(0.03) \\
Fe\,XX & 0.96 & 0.09(0.02) \\
Ne\,X & 1.022 & 0.46(0.02) \\
Fe??/Ne\,IX & 1.10 & 0.20(0.01) \\
Fe\,XX/Ne\,X & 1.22 & 0.08(0.01) \\
Mg\,XI & 1.33 & 0.28(0.01) \\
Mg\,XII & 1.47 & 0.29(0.02) \\
Mg\,XI & 1.60 & 0.06(0.01) \\
Al\,XIII & 1.73 & 0.08(0.01) \\
Si\,XIII & 1.85 & 0.30(0.02) \\
Si\,XIV & 2.00 & 0.15(0.02) \\
\hline
\multicolumn{2}{c}{Total SWCX (pn normalisation = 1.0)} & 12.58 (0.20) \\
\multicolumn{2}{c}{MOS1 Normalisation} & 0.80 (0.02) \\
\multicolumn{2}{c}{MOS2 Normalisation} & 0.92 (0.02) \\
\hline
\multicolumn{2}{c}{Reduced $\chi^{2}$ / Degrees of Freedom} & 1.17 / 1546 \\
\hline
\end{tabular}
\label{tabmodelparams}
\end{table}

\section{Viewing geometry and orientation of XMM-Newton}\label{secview}

The expected X-ray emissivity of SWCX emission from the solar wind
interaction with the magnetosheath can be estimated from the
integrated emission along the line of sight for the observer. 

The emissivity expected \citep{cravens2000} is given by the expression:

\begin{equation}
P_{^\chi} = \alpha\eta_{^{SW}}\eta_{^n}\langle g \rangle\ eV cm{^{-3}}s{^{-1}}
\label{eqnemiss}
\end{equation}

where $\alpha$ is a scale factor dependent on various aspects of the
charge exchange such as the interaction cross-section and the
abundances of the solar wind ions, $\eta_{^{SW}}$ is the density of
the solar wind protons, $\eta_{^n}$ is the density of the neutral
species and $\langle g \rangle$ is their relative velocity.

The flux is given by integrating along a particular line of sight:

\begin{equation}
F = \frac{1}{4\pi} \int_0^\infty P_{^\chi} ds\ ph\ cm{^{-2}}s{^{-1}}sr{^{-1}} 
\label{eqnflux}
\end{equation}

%%%%% Revise this section %%%%
\begin{figure}
  \centering
  \includegraphics[width=0.45\textwidth, bb=70 400 480 775]{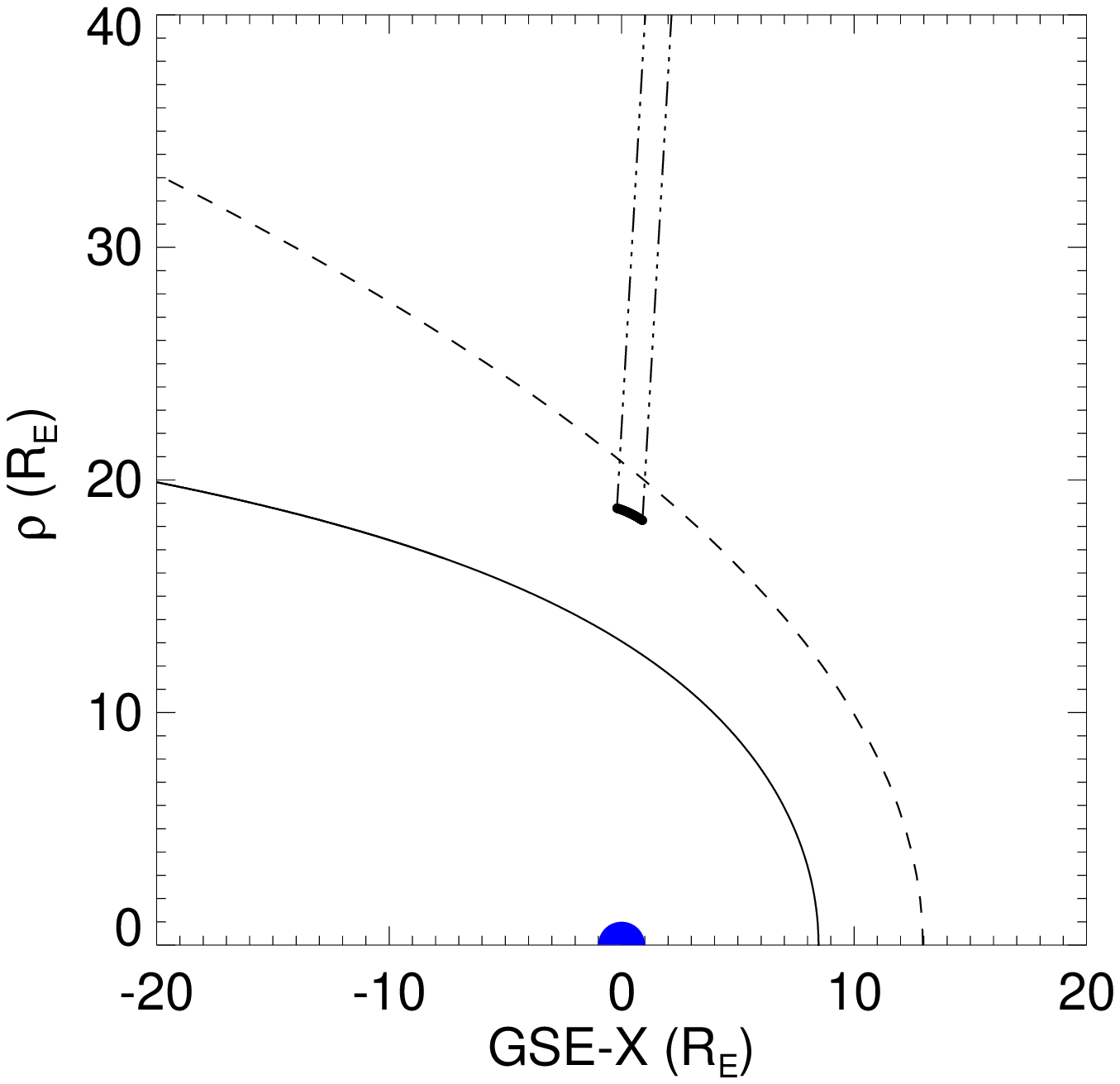}
  \includegraphics[width=0.45\textwidth, bb=70 400 480 775]{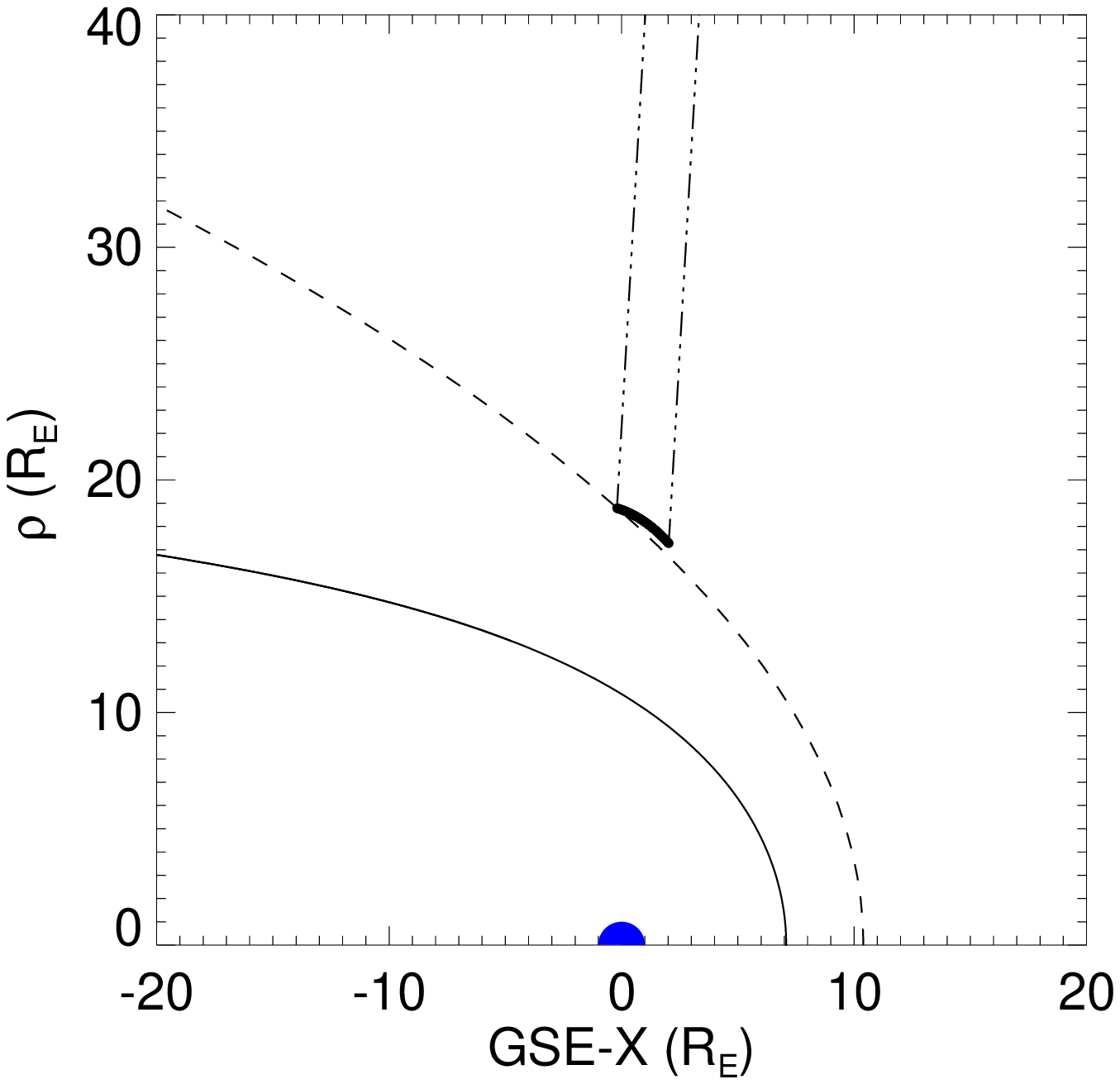}
  \includegraphics[width=0.45\textwidth, bb=70 400 480 775]{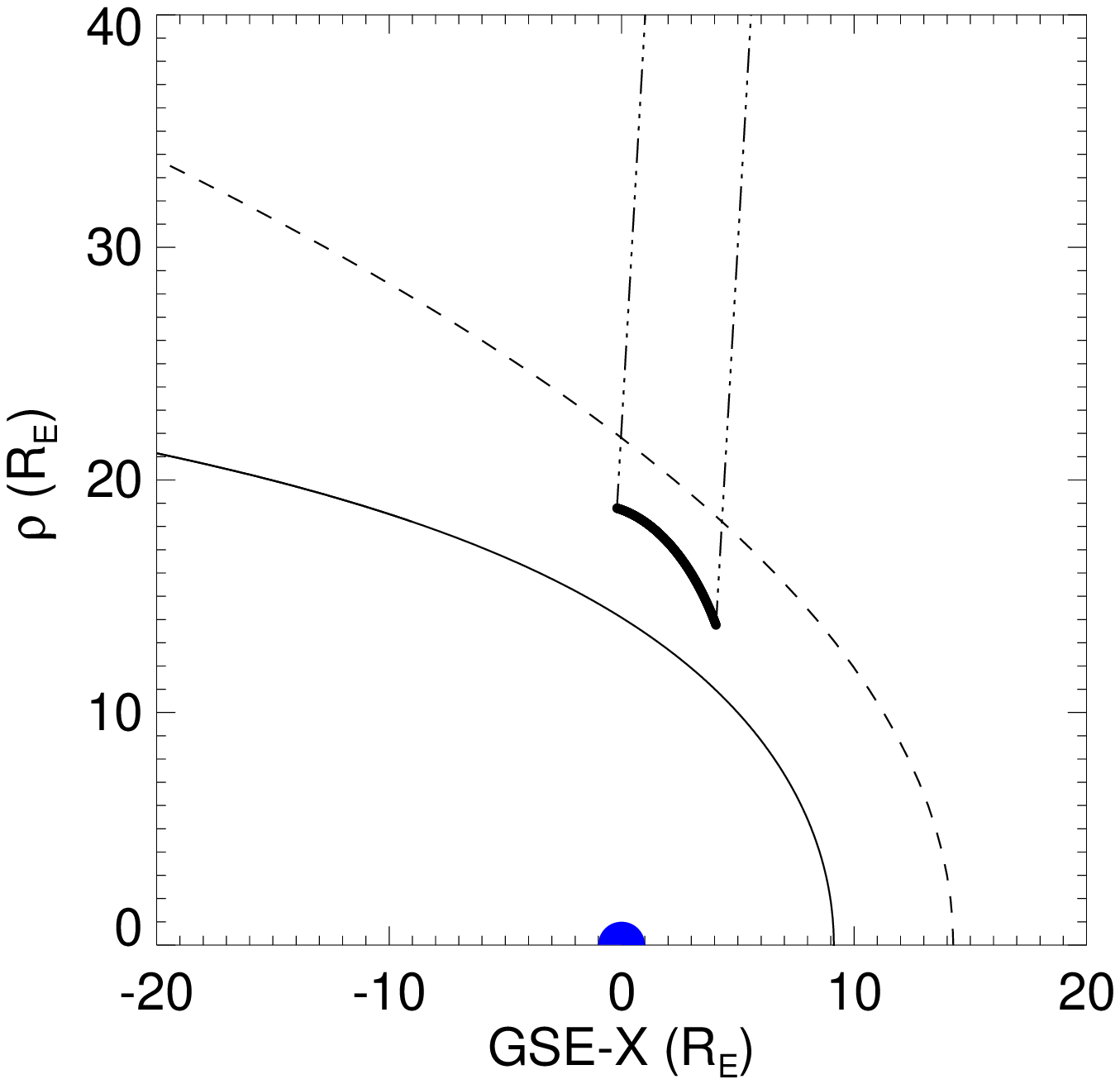}
  \caption{\ourc\ \xmm\ positions, in the ecliptic plane, progressing
    (top to bottom) from T=1.200805$\rm \times 10^{8}\,s$ to
    T=1.2009$\rm \times 10^{8}\,s$, through T=1.2010$\rm \times
    10^{8}\,s$ until T=1.2012$\rm \times 10^{8}\,s$. On the y-axis is
    $\rho$, as defined by $\rm \sqrt{GSE\mhyphen Y^{2}+GSE\mhyphen
      Z^{2}}$, plotted against the GSE-X coordinate. The solid and
    dashed lines give the locations of the magnetopause and bow shock
    at the end of each period. The dot-dot-dashed lines show the line
    of sight of \xmm.}
  \label{figxmmpossnaps}
\end{figure}

To assist in our analysis for this section, we took data from the
Solar Wind Experiment (SWE) instrument on board the spacecraft \wind\
\citep{ogilvie1995short} and data as measured by the SWEPAM instrument
of \ace.

In Figure \ref{figxmmpossnaps}, we plot the position of \xmm\ and the
magnetopause and bow shock boundaries and the line of sight during
\ourc. The line of sight pointed through the flanks of the
magnetosheath throughout the observation. \xmm\ crossed the bow shock
boundary as it moved along its orbit, and as the boundary position
changes in response to conditions in the solar wind. The magnetopause
and bow shock locations may vary dramatically, especially under
extreme solar wind conditions as we see in this case.  We used the
model of \citet{shue1998}, which takes the strength of one component
of the magnetic field ($B_{z}$) and the proton dynamic pressure as
input, to calculate the location of the magnetopause. The position of
the bow shock stand-off distance (the distance from the Earth to the
bow shock on the Earth-Sun line) is calculated using the solar wind
dynamic pressure \citep{khan1999}, and its shape is approximated using
a simple parabola (eccentricity 0.81). We plot the positions of \xmm\
as it moves from T=1.200805$\times 10^{8}\,s$ to T=1.2009$\times
10^{8}\,s$ (top), through T=1.2010$\times 10^{8}\,s$ (middle) until
T=1.2012$\times 10^{8}\,s$. The magnetopause and bow shock positions
are plotted for the end of each period. \xmm\ remains outside the
magnetopause for the entirety of the observation, but may cross the
bow shock boundary when the magnetosheath is compressed in response to
the solar wind (middle panel). Therefore the line of sight through the
magnetosheath region is short, but not zero, at various times during
\ourc.

%%%%% Modelling parameters %%%%%%%%%%%%%%%%%%%%%%%%%%%%%%%%%%%%%%%%
During \ourc\ the average solar proton density (level 2, hourly
averaged data from \ace), was measured as $\rm 13\,cm^{-3}$ and had an
average speed of $\rm 647\,km\,s^{-1}$. Exospheric neutral hydrogen
densities fall off as \erad$^{-3}$ and are normalised to a value of
$\rm 25\,cm^{-3}$ at a distance of 10\,\erad\ \citep{hodges1994,
  cravens2001}. Using Equation~\ref{eqnflux} and the solar wind
parameters above, we wished to estimate the expected X-ray emission
seen by \xmm\ at its average distance from Earth of 13.8\,\erad,
integrating out to 100\,\erad. By this method we compare the scale of
the emission as recorded by \xmm\ to a non-time resolved order of
magnitude estimation for the expected X-ray emission.

%%%% Need to justify why we used these Spreiter values - from the plot in
%%%% Figure 8, we should take values near the bow shock %%%%
The solar wind slows down and its density increases inside the
magnetosheath. The magnetosheath is the region between the bow shock
(where the solar wind slows down from supersonic to subsonic speeds)
and magnetopause (the boundary layer separating the plasma of the
solar wind and that of the Earth's magnetic field). In this
estimation, we base the starting point at the \textit{average}
distance of \xmm\ from Earth, so that the line of sight through the
magnetosheath region is short compared to the remaining line of sight
out to a maximum of 100\,\erad. We approximate a line of sight of
2.2\,\erad\ from the average position of \xmm\ to the bow shock
boundary, with the remainder of the line of sight intersecting
unperturbed solar wind. To approximate these changes, we scaled
hydrodynamical models of \citet{spreiter1966} (Kuntz, private
communication) to the magnetopause standoff distance of 8\,\erad\ and
extract factors for adjusting the solar wind parameters at the
relevant position within the magnetosheath. We increase the solar wind
density by a factor of 3.5 and reduce the solar wind speed by a factor
of 0.8 within the magnetosheath region only, and leave it undisturbed
outside the bow shock.

%%%%%%%%%%%%%%% Justify the value of alpha used %%%%%%%%%%%%%%%%%%
The value of $\alpha$ is dependent on the abundances of the ion
species contributing to the charge exchange process, along with the
cross-section and energy of each interaction in the energy band of
interest. For this estimation we consider only contributions from the
O\,VII and O\,VIII. SWCX emission is directly proportional to
$\alpha$, which is in turn proportional to the abundance of the ion
specie in question. We use the ratio of OVII to OVIII flux from our
spectral analysis in Section~\ref{secswcxspec}, the cross-sections
found in \citet{bodewits2007b} (assuming a solar wind with velocity
$\rm 600\,km\,s^{-1}$) and an oxygen to hydrogen ratio of 1/1780 as
given in \citet{schwadron2000} to derive an OVII to OVIII abundance
ratio of 0.085:0.915. We then calculate $\alpha$ for these two ion
species to be $\rm 2.3 \times 10^{-15}\,eV\,cm^{2}$. Although the
solar wind speeds during \ourc\ are more common of a fast solar wind
state, CMEs are enriched with high oxygen charge states and other
minor ions \citep{richardson2004}.

The total expected (oxygen band) X-ray emission along the line of
sight and for average solar wind conditions was estimated to be $\rm
9.5\,keV\,(1.5 \times 10^{-8}\,ergs) \,cm^{-2}\,s^{-1}\,sr^{-1}$. The
contribution from inside the magnetosheath is estimated to be $\rm
4.8\,keV\,(7.6 \times 10^{-9}\,ergs) \,cm^{-2}\,s^{-1}\,sr^{-1}$ which
represents approximately 50\% of the total. From our spectral
analysis, we observe a flux of $\sim \rm 18.9\,keV\,(\sim 3.02 \times
10^{-8}\,ergs) \,cm^{-2}\,s^{-1}\,sr^{-1}$ from the O\,VII and O\,VIII
emission lines, approximately 2 times greater than we estimate, but
which is consistent given the various assumptions as detailed
above. For example, the density of the plasma outside the bow shock
may be even higher than the values used in this calculation, due to
turbulent processes, localised density enhancements and/or the
anisotropic distribution of neutral atoms in the vicinity of the Earth
\citep{hodges1994}.

Higher levels of geocoronal SWCX emission would be expected had \xmm\
been observing a target that required a pointing vector that
intercepted the area of highest X-ray emission, namely around the
subsolar point, defined as the position of the magnetopause on the
sunward side of the Earth-Sun line \citep{robertson2003a,
  robertson2006}. The solar wind flux during \ourc\ was so high that
the magnetopause was pushed close to the Earth as a result of the
balance between the pressure of the solar wind and that of the Earth's
magnetic field. Therefore, only a very short line of sight of \xmm\
intersected the magnetosheath region for a large proportion of the
observation. The remainder of the line of sight intersected
undisturbed solar plasma. There was a sufficient density of neutral
donor atoms outside of the bow shock, interacting with a particularly
dense solar plasma, that a significant contribution to the SWCX signal
originated from this region, even though beyond the bow shock the
solar wind has not been slowed considerably or the density increased
as it would have been within the magnetosheath. The SWCX emission in
this case was emitted from both before and just beyond the bow shock
boundary. Clearly in cases where \xmm\ does not have an optimal view
through the magnetosheath, there is the possibility of detecting SWCX
emission from the local region.
%%%%%%%%%%%%%%%%%%%%%%%%%%%%%%%%%%%%%%%%%%%%%%%%%%%%%%%%%%%%%%%%%%

\begin{figure}
  \centering
  \includegraphics[width=0.4\textwidth, bb=100 380 500 700]{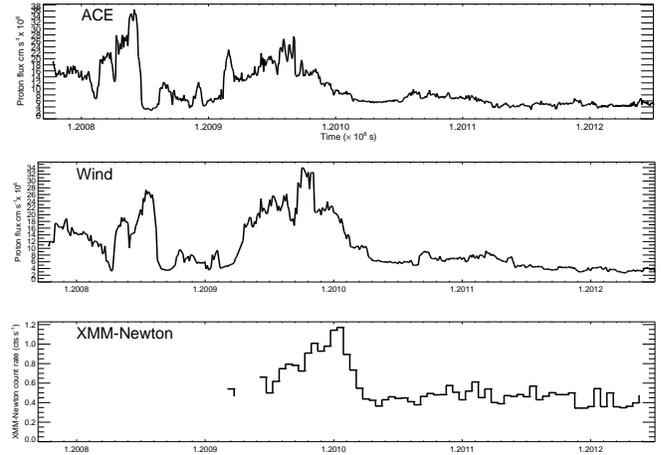}
  \caption{\ace\ and \wind\ solar proton densities plus \xmm\ combined
    EPIC instrument (0.5\,keV to 0.7\,keV, oxygen emission line band)
    lightcurve (panels top to bottom). The \xmm\ lightcurve has been
    flare-filtered using the method described in the text.}
  \label{figacewindxmmlc}
\end{figure}

In Figure~\ref{figacewindxmmlc} we plot the \ace\ and \wind\ proton
density lightcurves. We include on the plot the combined \xmm\ EPIC
instrument flare-filtered lightcurve, between 0.5\,keV and
0.7\,keV. The EPIC lightcurve shows the same general temporal
behaviour as the enhancements in solar proton density measured by both
\ace\ and \wind.

The offset in time between the signals at \ace\ and \wind\ is
explained by the separation between the solar wind monitoring
spacecraft. We are unable to determine the moment when the signal
first crossed into the field of view of \xmm, as the \xmm\ data had to
be heavily filtered for soft proton contamination at the beginning of
the observation. The shape of the lightcurve seen by \wind\ is not
exactly the same as that seen by \ace, so we infer that some evolution
of the CME may be occurring or that there are local inhomogeneities
within the CME wavefront although the bulk movement is fairly
constant. The shape of the \xmm\ lightcurve suggests some level of
averaging along the \xmm\ line of sight and it must be kept in mind
that the proton density is only a proxy for the ion composition of the
solar wind. From the positions of the spacecraft we are able to
ascertain that the CME wavefront extends at least 25\,\erad\ in the
GSE-Y direction. We have shown that SWCX emission is non-zero
throughout the \xmm\ observation, however we assume that the major
bulk of the CME has passed by a time at T=1.2012$\times 10^{8}\,s$. If
we take the start of the CME wavefront to be at approximately
T=1.20085$\times 10^{8}\,s$ travelling at an average speed of $\rm
647\,km\,s^{-1}$, the CME extends a minimum of 3500\,\erad\ in the
GSE-X direction.

\begin{figure}
  \centering
  \includegraphics[width=0.4\textwidth, bb=100 380 500 700]{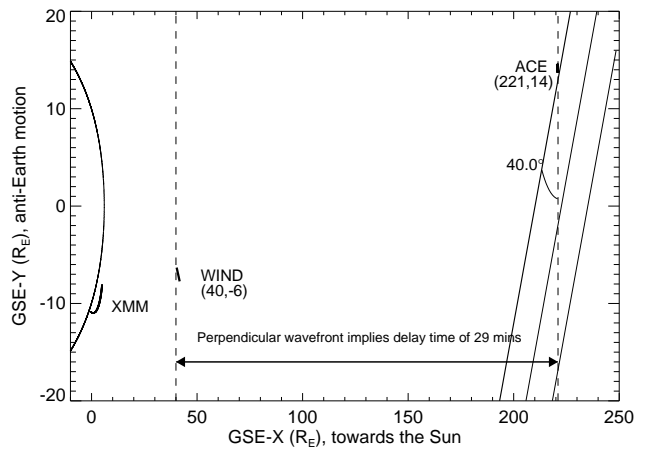}
  \caption{\xmm, \ace\ and \wind\ positions at the time of \ourc, in
    GSE-coordinates in the ecliptic plane. Positions in brackets are
    in \erad. The position of the magnetopause, as calculated using
    the \citet{shue1998} model, is shown for the start of \ourc. The
    dashed lines are used to aid visualisation and the wavefront is
    represented by the tilted solid lines.}
  \label{figacewindxmmpos}
\end{figure}

As the solar proton density lightcurves from both \ace\ and \wind\
showed the same shape we conclude that the same density enhancement
was received at both these solar wind monitors and subsequently
\xmm. We assume a planar wavefront for the enhancement, which is a
reasonable assumption at a distance of 1\,AU for a CME
\citep{zurbuchen}. Following a similar analysis to that of
\citet{collier2005} and \citet{collier1998} the orientation of the
passing wavefront could be derived using the delay between the signal
received at \ace\ and that received by \wind. Using a discrete
correlation function algorithm \citep{edelson1988} between the \ace\
and \wind\ proton density lightcurves (Figure~\ref{figacewindxmmlc}),
based on the period of the solar proton density enhancement between
T=1.2009$\times 10^{8}\,s$ and T=1.2011$\times 10^{8}\,s$, we find a
delay of 26$\pm$1 minutes from \ace\ to \wind. For a wave front
travelling perpendicular to the Earth-Sun line at the average speed
mentioned above, this results in a delay time of 29 minutes from \ace\
to \wind. The difference in delay times suggests that a tilted wave
front at approximately 40 degrees would have passed in the vicinity of
the Earth and \xmm. In Figure~\ref{figacewindxmmpos} we plot the
position of \ace, \wind\ and \xmm\ at T=1.2009$\times 10^{8}\,s$ of
\ourc.

\section{Discussion and conclusions}\label{secdiscuss}
We consider the possibility that the SWCX enhancement of \ourc\ is
linked to the CME event of the 19th October 2001 \citep{wang2005}. The
delay between the occurrence of the CME at the solar corona and its
arrival near Earth would be approximately two and a half
days. Increased solar proton fluxes were registered by both \ace\ and
\wind\ and therefore this plasma cloud would have passed in the
immediate vicinity of the Earth. It is not always the case that
enhancements in solar proton fluxes, and any accompanying highly
charged ions, are registered by increased incidents of soft proton
flaring or SWCX enhancements by \xmm. However, the arrival of the peak
of the low energy enhancement as seen by \xmm\ is consistent with the
delay expected as the feature passes in sequence from \ace\ to \wind,
on to a region intersected by the line of sight of \xmm.

We have shown that line emission from O\,VIII is very prominent and
dominates that of O\,VII, contrary to signatures of heliospheric SWCX
\citep{koutroumpa2009a}. We have also shown in our spectral analysis,
that the observed flux from SWCX emission is much greater than that
from a simple estimate of the expected emission, based on the
abundances of a slow solar wind. Also, mid-energy emission lines in
the regime 0.70\,keV to 2.00\,keV infer the presence of highly charged
states of iron, as is often seen in a CME \citep{zurbuchen, zhao}. We
see no significant compositional changes in the line emission over the
duration of the \xmm\ observation. In addition, we have observed
emission at 2.00\,keV from highly charged states of silicon, implying
a very high temperature plasma. A CME, rather than a steady state
solar wind, would explain the large enhancements, flux observed and
the richness of the spectrum as seen by \xmm. This case is the richest
spectrally of those examined by \citet{carter2008}.

CMEs have been used to explain the results of other X-ray observations
in the literature pertaining to the diffuse X-ray
background. \citet{henley2008} invoked a CME to explain differences
between results obtained from \xmm\ and \suzaku, when determining Halo
and Local Bubble X-ray spectra. They also observed emission from
Mg\,XI and Ne\,IX, although emission lines from oxygen were less
significant. They attribute this emission to a possible localised
enhancement in solar wind density crossing the line of sight of
\xmm. \citet{smith2005} attributed the anomalously high level of
O\,VIII seen in their observation of a nearby molecular cloud to SWCX,
and noted this was unlikely to be due to SWCX from a steady state
solar wind. Instead they conclude that their enhancement was due to
charge exchange from a CME and the interstellar medium, probably at a
distance of a few AU from the Sun, due to the depletion of neutral gas
available for charge exchange near the Sun. We eliminate the
possibility that the emission seen in \ourc\ is due to SWCX occurring
at the heliospheric boundary or at a large distance from
Earth. Short-term variations can occur for heliospheric SWCX,
especially if observing along the helium focusing cone
\citep{robertson2003a, robertson2003b}, but the pointing of \xmm\,
which does not insect the region of peak emission from this area,
argues against this case. In addition, the abundant emission line
spectrum and the variations in the fluxes of the major ions in the
spectrum which reflect the variations in solar proton flux support a
geocoronal occurrence of SWCX. We conclude that the SWCX interaction
we have observed occurs between ions from a CME and neutrals in the
exosphere of the Earth, at a relatively close distance to the Earth,
but not confined to the magnetosheath within the bow shock.

Although data regarding the ion states of the solar wind for the
period of \ourc\ from the solar wind monitors \ace\ and \wind\ are
sparse, we have been able to identify ions from a rich set of emission
lines from a passing CME. Not all CMEs detected by \ace\ will be
detected by \wind, or indeed intersect the line of sight of
\xmm. \xmm\ was not optimised to study the magnetosheath or near Earth
regions. However, we have shown that \xmm\ can be used to provide
additional compositional information of the solar wind plasma,
especially for the highest charge state ions, to that obtained by
upstream solar wind monitors, providing the observing geometries and
inclinations of the incoming wave fronts are favourable.

\section*{Acknowledgments}
We are grateful to Ina Robertson (University of Kansas), Michael
Collier (NASA, GSFC) and Mark Lester (University of Leicester) for
helpful discussions. We thank the anonymous referee for comments and
suggestions which have significantly enhanced this paper. This work
has been funded by the Science and Technology Facilities Council, U.K.

%\begin{thebibliography}{99}
\bibliographystyle{mn2e} % style
\bibliography{phd_gen_aa} % your references Yourfile.bib
%\end{thebibliography}

\label{lastpage}

\end{document}